# Revelations on Jupiter's Formation, Evolution and Interior: Challenges from Juno Results

Ravit Helled[1,] David J. Stevenson[2], Jonathan I. Lunine[3]
Scott J. Bolton[4], Nadine Nettelmann[5], Sushil Atreya[6]
Tristan Guillot[7], Burkhard Militzer[8,9], Yamila Miguel[10,11]
William B. Hubbard[12]

**Abstract** The Juno mission has revolutionized and challenged our understanding of Jupiter. As Juno transitioned to its extended mission, we review the major findings of Jupiter's internal structure relevant to understanding Jupiter's formation and evolution. Results from Juno's investigation of Jupiter's interior structure imply that the planet has compositional gradients and is accordingly non-adiabatic, with a complex internal structure. These new results imply that current models of Jupiter's formation and evolution require a revision. In this paper, we discuss potential formation and evolution paths that can lead to an internal structure model consistent with Juno data, and the constraints they provide. We note that standard core accretion formation models, including the heavy-element enrichment during planetary growth is consistent with an interior that is inhomogeneous with composition gradients in its deep interior. However, such formation models typically predict that this region, which could be interpreted as a primordial dilute core, is confined to ∼10% of Jupiter's total mass. In contrast, structure models that fit Juno data imply that this region contains 30% of the mass or more. One way to explain the origin of this extended region is by invoking a relatively long (∼2 Myrs) formation phase where the growing planet accretes gas and planetesimals delaying the runaway gas accretion. This is not the same as the delay that appears in standard giant planet formation models because it involves additional accretion of solids in that period. However, both the possible new picture and the old picture are compatible with the formation scenario recently proposed to explain the separation of two meteoritic populations in the solar system. Alternatively, Jupiter's fuzzy core could be a result of a giant impact or convection post-formation. These novel scenarios require somewhat special and specific conditions. Clarity on the plausibility of such conditions could come from future high-resolution observations of planet-forming regions around other stars, from the observed and modeled architectures of extrasolar systems with giant planets, and future Juno data obtained during its extended mission.

**Keywords:** planets and satellites: interiors; planets and satellites: composition

_________________________

[1]Institute for Computational Science, University of Zurich, Winterthurerstr. 190, CH- 8057 Zurich, Switzerland

[2]Division of Geological and Planetary Sciences, California Institute of Technology, Pasadena, California 91125, USA

[3]Department of Astronomy, Cornell University, Ithaca, NY 14853, USA

[4]Southwest Research Institute, San Antonio, TX 78238, USA

[5]DLR Berlin, Institut für Planetenforschung, Rutherfordstr. 2, 12489 Berlin, Germany

[6]University of Michigan, Climate and Space Sciences and Engineering, Ann Arbor, MI 48109, USA

[7]Universite Cote d Azur, OCA, Lagrange CNRS, 06304 Nice, France

[8]Department of Earth and Planetary Science, University of California, Berkeley, CA 94720, USA

[9]Department of Astronomy, University of California, Berkeley, CA 94720, USA

[10]SRON Netherlands Institute for Space Research, Sorbonnelaan 2, NL-3584 CA Utrecht, The Netherlands

[11]Leiden Observatory, University of Leiden, Niels Bohrweg 2, 2333 CA Leiden, The Netherlands

[12]Lunar and Planetary Laboratory, University of Arizona, Tucson, AZ 85721, USA

# 1. Introduction

Jupiter is the largest planet in the Solar System and hence key to revealing the processes involved in gaseous and remnant disk formation. Because Jupiter's composition is primarily hydrogen (H) and helium (He), it must have formed early. Its large mass profoundly influenced the formation of our planetary system, in particular, the dynamical evolution of small objects, including those that contribute to the make-up of bodies far away from Jupiter, for example, Earth. In addition, Jupiter's composition can be used to constrain the conditions of the protoplanetary disk from which the solar system formed. As a result, some of the outstanding questions associated with the formation of our planetary system (and others) might be answered by understanding Jupiter better. Limited data up through the era of the Galileo mission's explorations of Jupiter have hampered detailed understanding of the formation history of the planet, its long-term evolution and current-state internal structure (e.g., Lunine et al., 2004; Bolton et al., 2010, Guillot et al 2004, Helled et al., 2014, Bolton et al, 2017b). In July 2016 the Juno mission began to orbit Jupiter with one of its primary goals to investigate Jupiter's formation and evolution through the characterization of its interior structure and atmospheric composition (Bolton et al., 2017b). Juno's investigation of Jupiter's interior is anchored in accurate measurements of both its gravitational and magnetic fields (e.g., Bolton et al., 2017ab, Folkner et al., 2017, Iess et al., 2018, Connerney et al., 2017). In this review we focus on Juno's gravitational field and their implications for our understanding of Jupiter's origin, evolution and internal structure.

Jupiter internal structure models that fit Juno's gravity data indicate that the planet is not homogeneously mixed and that its core is diluted from a pure composition of heavy elements, elements heavier than helium (e.g., Wahl et al., 2017; Nettelmann, 2017, Vazan et al., 2018, Debras & Chabrier, 2019). These updated models imply that Jupiter's core cannot be thought of as a pure heavy-element central region with a density/composition jump at the core-envelope-boundary. Instead, the core region is extended with a composition that may contain a similar amount of heavy elements by mass but is so diluted that hydrogen remains the most abundant material by number and might even dominate by mass. In other words, the "core" of Jupiter can have similar properties to a hydrogen-helium mixture though denser. This is possible because the nominal heavy-element core would constitute only a tiny fraction of the total mass (typically 5%).

There is no unique solution for Jupiter's internal structure and more than one density profile can satisfy all the observational constraints. Differences in the inferred structure could be a result of different assumed structures and/or boundary conditions, as well as different assumed equations of state (EoSs) and even the exact method that is used to derive the gravitational coefficients. Indeed, one is dealing with such small differences that it is possible to have a "good" EoS, one that is as good as the data and theory that define it, and yet still have significant differences between interior models with respect to the innermost region because the fraction of the total planetary mass under debate is so small and the gravity data are not well suited to discerning details of interior structure near the planetary center.

Nevertheless, all of the models listed above suggest that Jupiter's dilute core extending to several tens of percent of its total radius, perhaps even beyond half the radius. In addition, the models suggest that the heavy-element distribution within the planet is inhomogeneous, implying that it is not fully convective. The fact that Jupiter's structure models have become more complex also implies that the updated interior models include more free parameters, which naturally leads to more uncertainty in the inferred properties. In particular, the uncertainty in Jupiter's bulk composition increases because the region of inhomogeneity can also be a region of non-adiabaticity, violating the usual assumption of an adiabatic temperature profile throughout the region dominated by hydrogen and helium. Typically, an uncertainty in temperature of 10% corresponds to an uncertainty in pressure of ~1% (a figure that is larger towards the surface but smaller towards the center) and this means that temperature can be traded off with composition.

For example, one could imagine an increase in heavy elements to the extent of 10 $M_\oplus$ towards the center being offset by increasing the temperature there to from 20,000K to 40,000K. Non-adiabatic models indeed show this tendency (Leconte and Chabrier, 2012, Vazan et al., 2018). At the same time the planet must be Rayleigh-Taylor stable so there is a limit to the increase in temperature over the adiabatic value that can be tolerated (Vazan et al., 2018, Debras & Chabrier, 2019, Debras et al., 2021).

The fact that structure models based on Juno data can no longer make the (over-simplified) assumption that Jupiter is uniformly mixed, and fully adiabatic (i.e., fully convective) with a pure heavy-element compact core provides key information on Jupiter's interior linked to its formation and evolution history. This is because the three aspects of formation, evolution and structure are coupled: the formation process determines the primordial internal structure and thermal state. This determines the heat transport mechanism as well as the potential re-distribution of heavy elements and helium, and the planetary long-term evolution (contraction and cooling rate). The evolution then determines the internal structure of Jupiter today. It is therefore clear that in order to link Jupiter's current-state structure with its origin, a good understanding of the thermal and structural evolution is required. In the bulk of what follows we examine which formation and evolution scenarios could be consistent with current Jupiter internal structure models.

## 2. Jupiter's interior structure models

Interior models are constructed to represent the current structure in the interior of Jupiter that matches the gravity data. Structure model solutions correspond to the used equations of state, assumed type of temperature profile (adiabatic or not), and a temperature boundary condition at the observed atmosphere, typically at the 1 bar pressure-level. Jupiter's gravity data have improved radically with the Juno mission (Bolton et al. 2017a), which has provided even and odd gravity harmonics with unprecedented accuracy (Folkner et al. 2017; Iess et al., 2018; Durante et al., 2020). This improved accuracy in the gravitational moments requires more sophisticated interior structure models. One interpretation (at least) shows that the presence of a dilute core that extends to ~0.5$R_{Jup}$ allows interior models to match Juno measurements of the gravitational moments $J_{2n}$ (Wahl et al., 2017; perhaps more definitively in Militzer et al, 2021) as we discuss below.

Jupiter's density profiles as inferred by various structure models are shown in Fig. 1. Presented are profiles from Wahl et al. (2017) with a dilute core, a profile from Debras & Chabrier (2019) where Jupiter is found to have a large inner region with composition gradients, the density profile inferred by Vazan et al. (2018) from an evolution model that fit Jupiter's current properties (but not all the gravitational coefficients). Also shown are recent density profiles from Nettelmann et al. (2021) using the H-He EoS of Chabrier et al. (2019) for two assumed 1-bar temperatures (166 K and 180 K), and two models from the ensemble inferred by Miguel et al. (2021). These density profiles accurately reproduce Jupiter's gravitational field as measured by Juno accounting for the effect of the winds on the gravitational coefficients (see review by Kaspi et al. 2020 and references therein for details) with the exception of the Vazan et al. model which corresponds to Jupiter's current-state structure based on an evolution simulation.

As can be seen from the figure, despite some differences in the density profile due to different model assumptions and the different EoSs being used, the density profiles have striking similarities. This is expected as all models fit the Juno measured gravity field (Folkner et al., 2017, Iess et al., 2018). Figure 1 is useful for showing that the outer boundary of the enriched region, which could represent the dilute core boundary, is different for each model. It is often represented as a density jump. The different models predict a density discontinuity at different radii. There is no physical or model requirement for a density jump as opposed to a continuous change in composition over a broader radial range, though if one existed it would be diagnostic of how Jupiter forms and evolves. A statistical analysis of Jupiter's structure models assuming different assumptions points to an inhomogeneous composition where the inner part of the planet is more enriched in heavy elements than the atmosphere, which is already observed to be enriched (e.g., Miguel et al., 2022).

Figure 2 shows the heavy-element distribution in Jupiter as predicted by various structure models. Despite some differences, most models allow for (but do not require) the existence of a pure, small heavy-element core, with a mass ranging between 1-5 $M_\oplus$. The existence of such a pure compact heavy-element is consistent with formation models (e.g., Valletta & Helled, 2019, 2020, Brouwers & Ormel, 2020) as we discuss below. However, at present the existence of such a small core is poorly constrained by observations because gravity data, despite exquisite precision, is very insensitive to planetary properties near the center. Several structure models fit the Juno gravity data without a compact heavy-element core. This particular model suggests that the innermost part of the planet has a heavy-element mass fraction of ~20%. Interestingly, this model is consistent with updated formation/evolution models that show that due to the high temperatures of the young Jupiter, mixing is sufficiently strong to erode primordial composition gradients, leading to similar Z values (Müller et al., 2020).

**Figure 1. Left:** Jupiter's density profile as inferred from recent structure models. Several models indicate the existence of a density discontinuity at ∼ 50-60 % of the planetary total radius. This location could represent the size of a dilute core. The enclosed mass of the dilute core region is about 30% of the total mass, i.e., of the order of 100 $M_\oplus$ but poorly defined. At the same time, structure models with a small compact core cannot be excluded. **Right:** Density differences of the various structure models. The reference model is taken to be a Jupiter with n=1 polytrope.

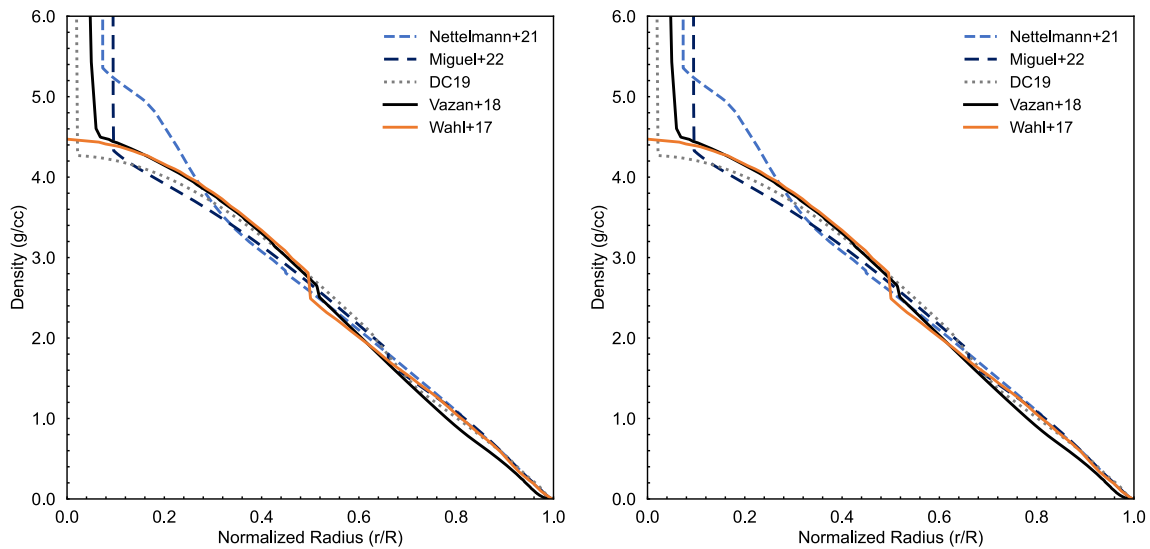

It is yet to be determined whether Juno's gravity data can be used to exclude the existence of a small inner compact core, which is composed of pure heavy elements. The possible elimination of a central pure core could be accommodated by changes further out, but the required changes seem to be *smaller* than the differences among the models that fit Juno gravity data. It is nonetheless of great value to use existing data to place constraints on the existence and mass of such a central "pure" core as it put certain constraints on constrain Jupiter's formation and evolution history. Outside this location, the heavy-element mass fraction Z is found to be as low as 20-30% and to decrease until a radius of 0.5-0.7 times the radius of Jupiter is reached; this is the location of the so-called dilute core outer boundary. We suggest that the question of whether Jupiter consists of a small pure heavy-element core should be further investigated: efforts should be made to identify the differences in the model assumptions and the thermodynamical properties of Jupiter for models with/without a small pure heavy-element core. This will allow us to understand how the requirement for a small compact core is compensated, and whether this is possible. If a detailed analysis on the requirement for having a small pure heavy-element core in Jupiter leads to the conclusion that such a compact core exists, this will usefully constrain Jupiter's origin and evolution, or alternatively, would require modifications in formation and evolution models.

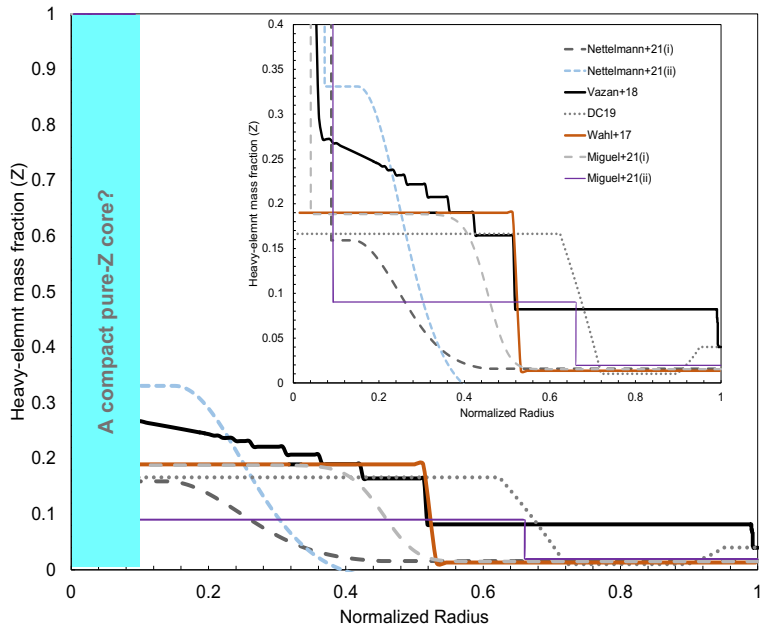

**Figure 2.** The heavy-element mass fraction in Jupiter as a function of normalized radius. The small panel zooms in to the region beyond the core. The various curves correspond to different published models as indicated in the figure. In some models, the innermost ~1% by mass (~10% by radius) of the planet is assumed to be a pure heavy-element region, i.e., a compact core while others can fit the gravity data without an inner pure heavy-element core. In all the models the heavy-element mass fraction in the planetary deep interior is moderate.

## 2.1 The Importance of the EoS of hydrogen for interior models

The EoS uncertainty is often thought of as being connected to the idea or computation of metallization in hydrogen. This, however, is not really the case since current Jupiter models neither compute or directly use this notion, except in the limited sense explained further below. In traditional discussions of giant planet models, much emphasis was placed on the metallization pressure for hydrogen and the possibly related exsolution pressure (helium rain formation pressure). The former, at least, is of less importance since there is no convincing evidence, either observationally or theoretically, of a first order phase transition in hydrogen metallization at the relevant pressure-temperature conditions in Jupiter's interior. In addition, hydrogen metallization at low temperatures does not necessary inform us on Jupiter since the molecular-metallic phase boundary (which can only exist below a critical temperature of perhaps ~2000K or less) is likely to be temperature dependent.

It is plausible to associate helium rain formation, where helium becomes immiscible in hydrogen and separates from it, which is a first order transition, albeit with a gradual change in helium composition with depth, with the onset of sufficient "free" electrons, and therefore the metallization of hydrogen, but this is also a theoretical construct. The exact conditions at which this separation occurs are still being investigated and incorporation this process in planetary conditions must reply on simulations (e.g., Morales et al., 2013, Schöttler & Redmer, 2018) and clearly more information on the EOSs of H and H-He is required. Note that traditional 3-layer structure models with a sharp "interface between the region with no helium rain and the region where limited solubility of helium is important are unrealistic since first of all, there is no reason to expect that interface between molecular and metallic hydrogen coincides with the one of He-poor and helium-rich regions, and second, a gradual change in helium is expected.

Currently there is no consensus on what constitutes the most realistic EoS for hydrogen alone, or H-He. Therefore, the existing Jupiter models often rely on different EoS and model assumptions. The uncertainties in the EoS, however, are important and not well understood. The uncertainties do not correspond to very low pressure (because we have experimental data) or at sufficiently high pressure (where the old theoretical idea of protons in an electron gas must be true, at least asymptotically) but are most profound at the intermediate region between ~0.1 to several Mbar. Unfortunately, in this pressure region the existing data are either too hot (shock waves) or too cold (diamond cell). Various studies have demonstrated the relevance of the EoS in the modelling of Jupiter's interior structure, specifically that the accuracy of interior models is directly linked with that of the EoS. The conclusion regarding Jupiter's inhomogeneous interior and the existence of the dilute/fuzzy core is a consequence of the recent improvements in the calculations of the EoS of hydrogen and helium at high temperatures and pressures (e.g., Becker et al., 2014, Militzer & Hubbard, 2013, Wahl et al., 2017, Nettelmann et al., 2021, Miguel et al., 2022). New calculations of the EoS of hydrogen predict a slightly higher density for hydrogen under jovian conditions (e.g., Militzer, 2013, Mazzola et al., 2018, Chabrier et al., 2019, Helled et al. 2020, Mazevet et al., 2020). As a result, for a given density, the resulting heavy-element mass fraction is smaller than in previous structure models that were based on the older EoSs such as SCVH (Saumon et al., 1995). In addition, Jupiter's $|J_4|$ value has been measured to be rather low. Both effects have led to structure models of Jupiter with a very low atmospheric metallicity. This result is inconsistent with the Galileo probe measurements (e.g., Debras & Chabrier, 2019 for details) and the new water measurements by Juno (Li et al. 2020; see Sec. 6 for caveats). In order to increase the envelope's metallicity to a reasonable value (i.e., at least solar), a dilute core must be assumed. It should be noted that in these new structure models of Jupiter the temperature profile is non-adiabatic, and the deep interior can be significantly hotter. Accretion and evolution models (e.g. Bodenheimer et al., 2018, Müller et al., 2020) show that the deepest region develops very high temperatures and is unable to cool efficiently because of the compositional gradient. Under such conditions more heavy elements can be incorporated for a given density.

A question that may arise is: *how robust is the currently-used hydrogen EoS that imply a non-adiabatic Jupiter with a dilute core?* While there is still an uncertainly in the EoS due to different functionals and numerical methods, multiple independent studies are in agreement at the few percent-level. This agreement is significant and a marked improvement on a decadal timescale, but it does not necessarily tell us that we have the right EoS because there is no experimental data at the conditions of greatest relevance for Jupiter. Moreover, these conditions are precisely those where hydrogen is not "simple": The molecules are breaking up, possibly forming chains of protons as well as individual protons and free electrons, and the internal degrees of freedom (so important for determining the isentrope) are no longer simple vibration and rotation. Shock wave experiments (in particular, Principle Hugoniots) are often used to assess the quality of the models but hydrogen is so compressible that these experiments correspond to higher temperatures than in the planet.

Other high-quality experimental data are available at low pressure and the first principles calculations should be accurate at very high pressure, so the potential source of disagreement between reality and theory lies in the intermediate regime where interpolation is necessary. The error needed to reconcile interior models with observations is only at the level of a few percent or so in density at a given pressure. Although the hydrogen EoS has been studied intensively there is still uncertainty (of the order of a few percent) linked to the EoS in the low-density range and the interpolation of the ab-initio simulations (e.g., Mazevet et al., 2020). As a result, Jupiter structure models are limited by the inherent uncertainty associated with the EoS used.

It should also be noted that atmospheric measurements by Galileo probe showed a depletion of He and Ne in Jupiter's envelope. This depletion is explained by a phase transition of He in Jupiter's interior (see discussion above and Stevenson & Salpeter, 1977) that was also found in numerical calculations of demixing of H-He mixtures (Morales et al. 2013) and recently also with experiments (Brygoo et al., 2021). The details of the H-He phase transition and how it might have affected the distribution of heavy elements both in the interior and the atmosphere of Jupiter are still largely unknown. This process is thought to be more important for Saturn, which indeed shows a larger depletion in helium. Helium rain can also affect the planetary evolution and lead to the formation of boundary layers that in turn would affect the heat transport. In addition, this process provides an additional energy source that "delays" the planetary contraction (e.g., Mankovitz & Fortney, 2020). Note that current Jupiter models assume that the average He abundance in the protosolar component of Jupiter is ~27.2% by mass. This is actually a model number (based on models of the Sun) and not observationally verified for Jupiter.

The presence of other constituents beyond He complicates the EoS and the phase diagram but this is probably less important than the concerns about pure hydrogen, given their lower abundances and the demonstrated near accuracy of volume additivity. Nevertheless, the additional elements add other source of complexity to structure models. Finally, current modeling on possible variations in EoS is incomplete but they suggest that they could affect both the issue of a central inner core and the heavy-element enrichment of the envelope, depending on pressure(s) at which the changes are made. Clearly experiments could advance our understanding of materials at planetary conditions independent of the inherent uncertainties in the theory.

## 3. Jupiter's Formation models

Our understanding of how gas giant planets form is still incomplete (see Helled et al., 2014 for review). There are two main scenarios for giant planet formation known as "core accretion" (e.g., Bodenheimer & Pollack, 1986, Pollack et al., 1996; Alibert et al., 2005) and "disk instability" (e.g., Boss 1997, 2000, Mayer et al. 2002, Durisen et al., 2007). In the core accretion model, the formation of a giant planet begins with the buildup pf a heavy-element core followed by gas accretion. In the disk instability model, on the other hand, giant planets form as a result of a local gravitational instability in protoplanetary disks which can be followed by accretion of heavy elements, and even core formation. These two formation scenarios typically operate at different conditions and timescales, and also often (but not always) lead to different predictions regarding the planetary final structure and composition (Helled et al., 2014, Bolton et al. 2017b). In this review we focus on the core accretion model, which seems to be the favorable scenario for Jupiter's formation.

In the core accretion model, a giant planet is formed via three main phases. The earliest stage of giant planet growth, referred to here as phase-1, involves accretion of nearly pure solid material and may take as little as a few hundred thousand years. The outer layer of this heavy element core may vaporize and become supercritical (i.e., hotter than the critical temperature) because of the energy of accretion (an unavoidable feature that is absent in the original model proposed by Pollack et al., 1996), once the core has reached a few $M_\oplus$. The overlying gas envelope is then small in mass. Phase-2, which is also known as the "attached phase", in which the growing planet is embedded in the proto-planetary disk, is dominated by slow accretion, mediated by the need to eliminate the accretional heating. The formation of composition gradients during planetary growth is expected to occur at the end of this phase, when the H-He accretion rate has become comparable to that of the heavy elements-- i.e., up to the beginning of runaway gas accretion (phase-3). This is also approximately when the H-He mass becomes larger than the heavy-element mass—a point known as the crossover mass. In the runaway, gas accretion is so large that the disk cannot supply the mass of gas required by the contraction of the planet. The gas accretion rate during runaway and the mechanism for termination of rapid gas accretion are not well determined and are topics of intense investigations. For example, in a disk of low-viscosity, gap opening can regulate the supply of disk gas during the runaway phase (e.g. Ginzburg & Sari, 2018), while in the high-viscosity case the gap that forms is shallow gap and the gas accretion rate is higher (e.g. Tanigawa & Tanaka, 2016). In any case, during this last stage of gas accretion, the planet gains most of its mass, and the composition becomes H-He dominated. There may also be a later phase of a small amount (e.g. a few $M_\oplus$) of solids accreted to explain the observed atmospheric enrichment, but this is still being investigated (Shibata & Helled, 2022).

In the classical giant planet formation models, in order to simplify the numerical calculations, it was assumed that all the heavy elements reach the center while the envelope is composed of solely H-He, resulting in a simple internal structure of a nearly pure heavy-element core surrounded by a H-He envelope (e.g., Pollack et al., 1996). However, more recent formation models that follow the heavy- element distribution during the formation process show that once the core mass reaches value of just a couple of Earth masses and is surrounded by a small envelope, the solids (heavy elements) tend to dissolve in the envelope instead of reaching the core. This result is insensitive to the assumption of sizes of the accreted heavy-elements and is found to be valid for both pebble and planetesimal accretion (e.g., Iaroslavitz & Podolak, 2007, Hori & Ikoma, 2011, Brouwers et al., 2018, Valletta & Helled, 2019).

Recent giant planet formation models that follow the heavy-element deposition in the atmosphere during planetary growth show that Jupiter's primordial structure is characterized by a deep interior highly enriched with heavy elements (e.g., Helled & Stevenson, 2017, Bodenheimer et al., 2018). The vaporization of the heavy-element core early in phase-1 described above means that heavy elements can remain in the planetary envelope instead of reaching the core. However, the atmosphere can become supersaturated in heavy elements, leading to rain-out and even more central concentration than one would infer simply by looking at the accretion fluxes of solids and gas (Bodenheimer et al, 2018, Valletta & Helled, 2019, 2020, Stevenson et al, 2022). This implies that Jupiter's primordial structure includes a deep interior that is highly enriched with heavy elements, with no sharp transition between the small heavy-element core and the inner envelope (e.g., Lozovsky et al., 2017, Helled & Stevenson, 2017, Valletta & Helled, 2019, 2020, Ormel et al. 2021 and references therein). In the absence of later redistribution, the *Z(r)* depicted in Figure 2 for present-day Jupiter tells us *Z(m)* where *m* is the mass enclosed within radius *r* at the time of accretion. The radial coordinate has changed because the planet contracts greatly in the tens of millions of years immediately after cessation of accretion.

It was suggested by Helled & Stevenson (2017) that the nature of the gradient in *Z(m)* is linked to the ratio between the heavy-element and H-He accretion rates. Detailed formation models (Valletta & Helled, 2020) confirm the relationship between the obtained Z(m) and the relative fluxes before runaway gas accretion takes place. The exact relationship may be more complex and likely also affected by the vapor pressures of the accreted solids (e.g., Bodenheimer et al, 2018; Stevenson et al, 2022). Nevertheless, assuming no significant mixing occurs post-formation, then the heavy-element profile is linked to the accretion rates via the following approximation (Helled & Stevenson, 2017):

$$Z(m) = \frac{dM_Z/dt}{\left(dM_{H-He}/dt + dM_Z/dt\right)}, \qquad (1)$$

with the understanding that the fluxes of heavy elements (*dM_Z/dt*) and hydrogen-helium (*dM_{H-He}/dt*) on the right hand side of the equation are evaluated at the time when the accreted mass is *m*. This is valid under the following conditions: as heavy elements are accreted, a specific amount of gas is accreted simultaneously in order to maintain hydrostatic equilibrium between the planet and the nebula. Provided the solids vaporize high up and stay high up, they are accreted at roughly the same radial location as the newly accreted gas.

A possible objection is that incoming solids do not necessarily end up near the radius at which they evaporate upon impact. Some may sink deeper as droplets because of over-saturation; (Bodenheimer et al. 2018, Stevenson, 2022). The disposition of the solids depends on the size of the accreted bodies, which range all the way from 10-cm-scale pebbles, through planetesimals of a few to a hundred kilometers in radius, up to giant impacts that mechanically stir the entire body therefore the exact distribution depends on the model assumptions. Nevertheless, recent giant planet formation simulations confirm the validity of this approximation for the innermost regions of the planet (Valletta & Helled, 2020).

Standard formation models of Jupiter suggest that the runaway gas accretion occurs when the protoplanetary mass is of the order of a few tens of $M_\oplus$. Therefore, in these cases the composition gradients predicted from formation models are always limited to the innermost region of the planet, about 10% of its mass. During the last stage when most of the mass is accreted the total heavy-element mass is expected to be rather low (e.g., Shibata & Ikoma, 2019). Therefore, standard formation models provide a path to form Jupiter with a small pure heavy-element compact core surrounded by composition gradients that can be viewed as a dilute core. However, the dilute core arrived at by standard formation models may not look like that required by structure models. The qualitative difference between the heavy element distribution inferred by formation models to those inferred from structure models is demonstrated in Figure 3. The metallicity of the molecular envelope is not well determined, as indicated by the gray box. In fact, both formation and structure models predict a metal-poor envelope which is at odds with the measurements of the Galileo probe. Formation models suggest that planetesimal accretion during the late stages of gas accretion is not very likely, unless Jupiter migrated significantly (e.g., Shibata et al., 2021, Shibata & Helled, 2022). Mechanisms that can lead to atmosphere enrichment include the erosion of composition gradients within Jupiter or a giant impact (discussed below). Interestingly, structure models that fit Juno gravity data tend to predict a very low (or even negative!) envelope metallicity. It is possible that future modifications of the H EOSs would predict a lower density for hydrogen at the relevant P-T regime, which in turn allows a higher metallicity. At the moment this is speculative. Finally, it is possible that the atmospheric metallicity of the uppermost part of Jupiter's atmosphere does not represent the bulk of the molecular envelop that is expected to reach down to pressure of ~ 1 Mbar and that the measured enrichment is a result of a recent pollution. However, the stability of such an enriched layer against overturn is questionable. This important topic should be investigated in detail in the future; it could also have consequences for the other giant planets in the solar system as well as giant exoplanets.

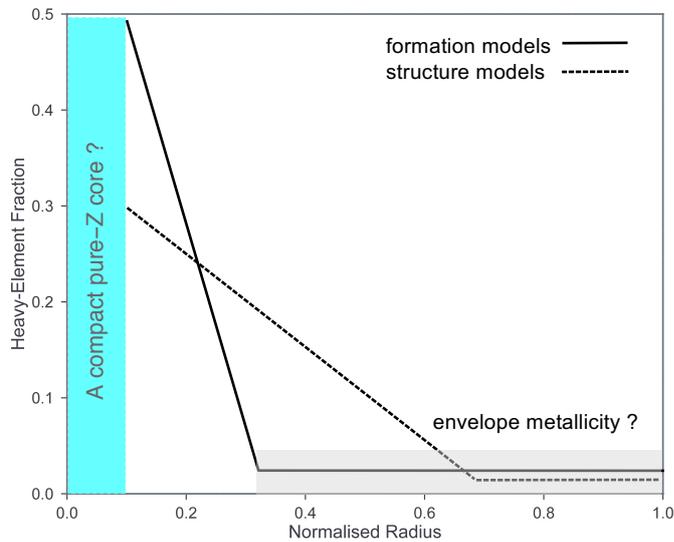

**Figure 3.** A sketch of the heavy-element mass fraction as predicted by structure (dotted) and formation (solid) models. It is clear that although both models suggest the existence of a composition gradient, the gradient inferred from structure model is much more extended and the Z value is significantly lower. The light grey box correspond to the profound uncertainty in the envelope metallicity of Jupiter in both formation and structure models.

It should be noted that additional key information on Jupiter's origin comes from the pattern of heavy element abundances in Jupiter's outer envelope. The basic feature of the abundance pattern is that, excluding helium and neon which are likely involved in phase separation associated with the transition of hydrogen to the metallic phase, the enrichment relative to solar is consistent with a uniform factor of between 3-5 (e.g., Atreya et al., 2019). Although the abundance of water is crucial because oxygen is potentially the most important heavy element, its abundance in Jupiter remains uncertain. First, the Juno derived results, while impressively deep below the formal condensation level for water clouds, apply only to a single latitude near the equator (Li et al., 2020). Second, the error bars for water are sufficiently large that values as low as sub-solar are possible. The low value measured by the Galileo probe has long been argued to be a localized dynamical effect at the Galileo entry site (e.g., Wong et al., 2004). However, it is still not possible to rule it out as a global depletion. Better determining the water abundance in Jupiter (Li et al., 2020) is essential if we are to constrain different formation and evolution scenarios (e.g., Helled & Lunine, 2014, Bolton et al., 2010, 2017b).

Assuming the water abundance in Jupiter's atmosphere is between 1/2 and three times solar, one might ask what is required to achieve this depletion relative to the rest of the volatile pattern. One way is to alter the bulk C/O value in the protoplanetary disk so as to have less water ice in the disk compared to other condensables, but other evidence for such variations in the elemental ratio in the disk is lacking in primitive bodies. Also, while variations in the C/O value can occur in the gas phase of the protoplanetary disk as a function of time and location due to condensation, these do not directly translate into similar enhancements in C/O in the planetesimals. An alternative model considers the difference in behavior between planetesimals and smaller pebbles in a gaseous disk. Pebbles are expected to move radially a significant distance through the disk (Monga and Desch, 2015, Mousis et al., 2019). Amorphous ice in cold pebbles that cross the temperature threshold for conversion to crystalline ice will release most of their volatiles into the gas phase, even if the resulting crystalline ice were in the form of clathrate hydrate. As the proto-Jupiter grows and reaches the pebble isolation mass, the pebbles are no longer accreted efficiently but are trapped in the pressure bump, and the species more volatile than water ice will enrich the envelope of growing Jupiter through the gas phase, while leaving the water behind as ice in the pebbles. In this model, the Jovian envelope volatile pattern except for O would look like that of the enriched gas in the disk, while the amount of water accreted would depend on the temperature, the mass of water in pebbles vs planetesimals, and Jupiter's formation location (Mousis et al., 2019). There are still many open questions regarding the expected enrichment of Jupiter and this topic is currently being investigated intensively.

Overall, the possibility that the heavy element abundance is not constant with depth in the envelope, and that catastrophic events like a giant impact might have scrambled the volatile pattern in the violent rearrangement of the envelope makes the links between atmospheric and bulk composition, and origin and internal structure even more challenging. Finally, the characteristics of the Galilean satellites could also be used to further constrain formation models. The Juno extended mission is expected to provide more intriguing measurements of Jupiter's atmospheric composition and its satellite system.

## 3.1 Alternative paths for explaining Jupiter's dilute core

### 3.1.1. An Alternative Formation History: extended phase-2 via planetesimal accretion

Recent constraints on the formation timescale of Jupiter were claimed from isotopic anomalies found in iron meteorites (Kruijer et al., 2017). It was suggested that the carbonaceous and non-carbonaceous chondrites were separated from each other between ~1 and ~3-4 Myr after calcium-aluminium-inclusions (CAIs) formation. More specifically, it was suggested that proto-Jupiter reached a pebble isolation mass (Kruijer et al., 2017) of ~ 20 $M_\oplus$ at time 1 Myr. Subsequently, planetesimals were formed both inward and outward of the orbit of proto-Jupiter, until ~ 2 Myr for the non-carbonaceous chondrites and until ~3-4 Myr for the carbonaceous chondrites. At that time, Jupiter became massive enough (~50 $M_\oplus$) to scatter planetesimals and reconnects the two reservoirs. Note that the overall timescale is only marginally consistent with the time over which the protoplanetary disk gas is believed to be present (~3 Myr), based on astrophysical observations of comparable systems elsewhere (e.g., Williams and Cieza, 2011).

This formation scenario was investigated in detail by Alibert et al. (2018) where it was shown that this time constraint can be fulfilled with a heavy-element accretion rate of $\sim 10^{-6}$–$10^{-5} M_\oplus$/yr is needed to prevent the onset of runaway gas accretion before time ~3 Myr. Since proto-Jupiter reached the pebble isolation mass at time ~1 Myr, after this time pebbles could not be accreted by the planet, and hence, the heat source during that period must come from planetesimal accretion. It was also shown that the core mass is expected to be between 6 and 15 $M_\oplus$. Finally, it was confirmed by Venturini & Helled (2019) that indeed an initial phase of core formation is expected to be dominated by pebble accretion, followed by a second stage of planetesimal accretion, in a model whose aim was to form a Jupiter with a heavy-element total mass of 20–40 $M_\oplus$.

This modified formation scenario that includes accretion of both pebbles and planetesimals suggests that Jupiter grows via three main (modified) phases: *Phase-1*: core formation dominated by pebble accretion, with a high heavy-element accretion rate and negligible accretion of H-He. *Phase-2*: Jupiter is growing by accreting both planetesimals and H-He gas from the disk. The accretion rate of planetesimals is sufficient to prevent the planet from cooling down efficiently and reaching rapid gas accretion. *Phase-3*: Jupiter is massive enough (~ 100 $M_\oplus$) to reach the detached phase and runaway gas accretion takes place. There is of course a limit set by the astrophysical observations suggesting that the nebula is only present for ~3Myrs, presumably the formation time of all of Jupiter, except for a possible late addition of solids.

This is not conceptually different from standard giant planet formation models but leads to a different outcome because the modified *Phase-2* has a higher accretion of planetesimals that delays the cooling (Kelvin-Helmholtz contraction) needed to allow the onset of the runaway gas accretion. Runaway begins only when proto-Jupiter has already reached a mass of ~ 100 $M_\oplus$. This formation scenario can explain a dilute region outside the central region that conventional models predict and might go at least part way toward explaining Jupiter's dilute core (see Fig. 8 of Venturini & Helled, 2019). Such a formation scenario can therefore explain Jupiter's interior. Interestingly, the *Phase-2* associated with planetesimal accretion leads to accretion of material with $Z \sim 0.3$ which is consistent with recent structure models described in section 2.

Although it is not necessary to attribute the isolation of meteorite populations to the formation of Jupiter (Kruijer et al, 2017), the alternative accretion story fits with a possible interpretation of those meteorite data. In order to separate the meteoritic populations runaway must have started later on, via a state of planetesimal formation which delayed runaway gas accretion and at the same time provides the required metallicity and stopped at the right mass. It should be noted, however, that the transition to *phase-3* at this mass corresponds to an assumed solid surface density in the disk of 25 $g\ cm^{-2}$. This value is somewhat high but consistent with values inferred from models of planetesimal formation (e.g., Drażkowska et al. 2016). It is also important to note that there are large uncertainties in the sequence of events leading to the accretion of Jupiter so that such a formation scenario, although appealing, is clearly non-unique.

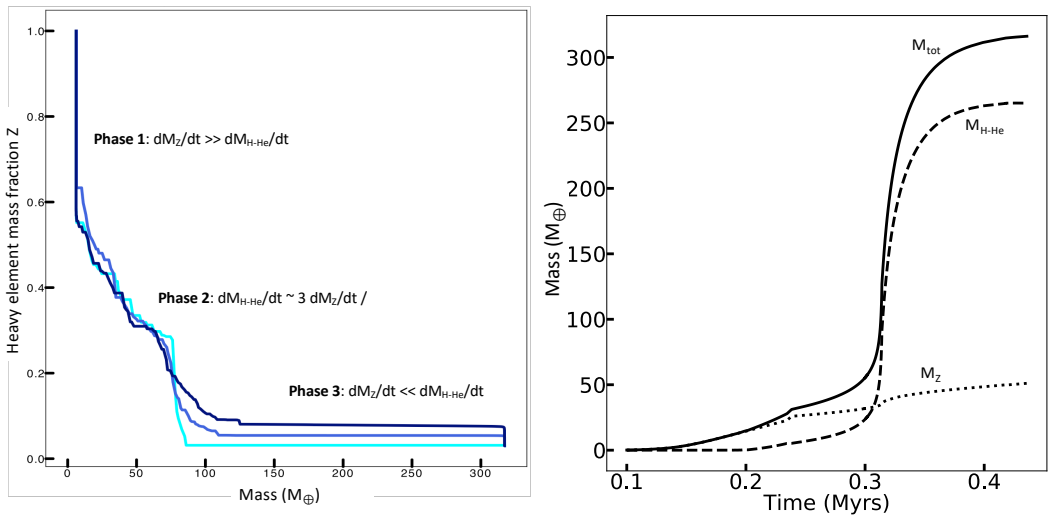

**Figure 4. Left:** The heavy-element mass fraction accreted at different stages of Jupiter's formation assuming the core is formed by pebble accretion followed by planetesimal accretion and finally rapid gas accretion (e.g., Venturini & Helled, 2019, Valletta & Helled, 2020). The three different curves correspond to different assumed solid surface densities. The profile represents the primordial heavy-element distribution in the deep interior.
**Right:** Planetary mass as a function of time. The dotted and dashed line correspond to the heavy-element and H-He mass, respectively.

As discussed above and shown in Eq. 1, there is a relation between the heavy-element profile and the accretion rates. This implies that by observing the heavy-element distribution in Jupiter today, we can put limits on its accretion rates, assuming that no significant mixing had occurred, and therefore on a key aspect of its formation history. Based on this assumption, if Jupiter has a small compact heavy-element core it represents phase-1 at which $dM_Z/dt >> dM_{H-He}/dt$. Then, if the innermost 30% of the mass is dominated by $Z \sim 0.3$, it would suggest that $[(dM_{H-He}/dt)/(dM_Z/dt)] \sim 2.3$, and during the last phase, since Z is nearly solar, it suggests that $dM_{H-He}/dt >> dM_Z/dt$, and the planetary growth is dominated by the H-He accretion rate, i.e., $(dM_{H-He}/dt)$.

This idea is demonstrated in Figure 4 that shows a sketch of the predicted heavy-element mass fraction as a function of planetary mass from the formation scenario discussed above.

An alternative formation model of Jupiter with a longer formation timescale due to an extended *phase-2* could also help to resolve the "fine tuning" issue of the formation of Uranus and Neptune (Alibert et al., 2018). Conventionally, Saturn is viewed as being like Jupiter except that the runaway phase was truncated at a much lower mass, permitting an envelope that is capable of being more readily enriched to the observed highly non-solar value. Alternatively, one could consider the possibility, based on the analysis above, that Saturn never entered into the runaway gas accretion phase and the enriched envelope is then testament to the accretion of planetesimals during the second prolonged phase of accretion (before 100 $M_\oplus$ is reached). It is unclear in this picture whether we should connect the observed threefold accretion of Jupiter's envelope with the possible tenfold enrichment of Saturn's envelope, notwithstanding the hint that the enrichment is in inverse proportion to the final mass.

Recent models that use both gravity and seismology Cassini data to constrain Saturn's internal structure (Movshovitz et al., 2019; Mankovich & Fuller, 2021) indicate that like Jupiter, Saturn also consists of a dilute core or composition gradients in its deep interior. The fact that Saturn is also expected to have a fuzzy core supports the scenario under which Saturn never reached runaway gas accretion. Unfortunately, we lack any information on the heavy noble gases in the Saturnian envelope, essential to constraining planetesimals composition with the same fidelity that the Galileo Probe measurements do for Jupiter.

To summarize, a modified formation scenario for Jupiter with an extended phase-2 is a possible reconciliation of Jupiter's dilute core and the meteoritic data, constraints that are completely independent from each other. We suggest that this provides a compelling reason for considering this formation path as a possible model for Jupiter's formation. Nevertheless, it remains to model the long-term evolution of the planets under this formation scenario in order to investigate whether the predicted heavy-element distribution profile is sustainable for timescales on the order of a few $10^9$ yrs.

### 3.1.2. A Giant impact post-formation

Another potential explanation for Jupiter's fuzzy core is a giant impact. The idea of giant impacts shaping the compositions and internal structures of Jupiter and Saturn was proposed before Juno arrived at Jupiter (e.g., Li et al., 2010). Recently, the giant impact scenario has been invoked to explain the origin of Jupiter's dilute core (Liu et al., 2019). The idea is that the young Jupiter shortly after its formation suffered a nearly head-on collision with a massive impactor that provided the energy necessary to disrupt the primordial compact core and mix the heavy elements into the envelope, as summarized in Figure 5 from Liu et al (2019). The post impact thermal evolution was modeled, and it was found that if the post-impact Jupiter is cold enough, significant mixing of the interior could be avoided, sustaining the fuzzy core until today (see Liu et al. (2019) and Müller et al., 2020 for details).

In order to dilute the core as a result of an impact, very specific conditions are required. The most favorable case is an impactor that is massive (e.g., 10 $M_\oplus$ or more; mostly heavy elements) with a small impact parameter. A high velocity at infinity is also desirable. It is questionable whether such a high velocity is realistic.

By considering both energy and angular momentum conservation, it follows that

$$(b/R)^2 = [v_t^2/(v_r^2+v_t^2)][1+2GM/Rv_\infty^2], \qquad (2)$$

where $b$ is the impact parameter, $v_\infty$ is the encounter velocity ("velocity at infinity"), $v_r$ and $v_t$ are the radial and transverse velocities of the projectile at impact at radius $R$, and $M$ is the target mass. For simplicity, the two-body approximation is used in which the projectile is much less massive than the target, but this is not a critical assumption. "Radial" means directed towards the center of the target.

Just as with throwing a dart at a dartboard of radius $b$, the probability of having an impact parameter less than or equal to $b$ scales as the area, $b^2$. It follows that if we define a "head on collision" as one where the tangential velocity is at most 10% of the radial velocity, then it happens in only about 1% of all collisions. Close encounters are even more common than oblique collisions and problematic if Jupiter had a disk. Notice that the expected low velocity at infinity can greatly increase the likelihood of a collision but does not favor head-on collisions.

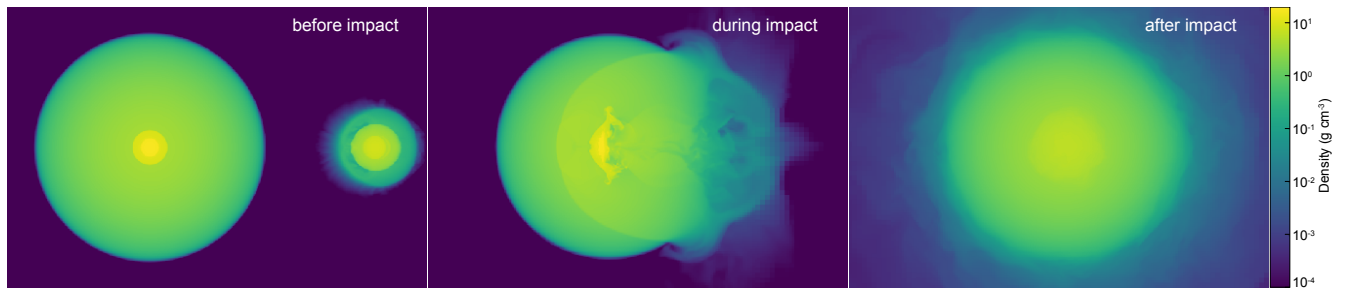

**Figure 5**. Snapshots of the density distribution in Jupiter before, during and after the giant impact as inferred by Liu et al. (2019). Jupiter pre-impact is assumed to have a 10 $M_\oplus$ compact core and the impactor is assumed to have a mass of 10 $M_\oplus$. The impact leads to mixing which decreases Jupiter's core density by a factor of three and leads to an extended dilute core.

In planet formation models, giant planet embryos can easily exceed the number of final giant planets (e.g., Levison et al., 2010) so merger events are permissible; future simulations are needed to evaluate the implications of multiple planetary embryos passing through Jupiter's location at various impact parameters. Giant impacts occurring during earlier stages of planetary formation might more readily mix the compact core with a smaller impactor and require less particular impact geometries. It is difficult to get enthusiastic about such a head-on catastrophic event for Jupiter if the same must be invoked for Saturn. But the giant impact scenario cannot be excluded and must remain in contention as an explanation for fuzzy cores.

## 4. Jupiter's Evolution

### 4.1. Core Erosion: Energetic considerations

Jupiter convects heat from its interior and therefore functions as a heat engine. This engine is capable of doing work; in particular, it can carry heavy material upward at the expense of light material. This gravitational work is limited by the buoyancy flux of the thermal convection. Put simply, an element of fluid may go up because it has thermal buoyancy that is larger than the magnitude of the (negative) buoyancy associated with the upward redistribution of heavy material. The problem with this physically appealing picture, the basis of early arguments offered in Stevenson (1985) and more quantitatively in Guillot et al. (2004), is that they depend on poorly understood aspects of convective mixing, especially the non-local character of convection. This is best illustrated by considering two possible core erosion stories, in which the first ("global") story predicts that any plausible core could be destroyed (i.e., Jupiter should be homogeneous) whereas the second ("local") story says that core erosion does not disperse an originally nearly pure core. The reality is surely in between, and so far existing models/experiments are insufficient to place useful constraints on core erosion. Most plausibly, as discussed below, it is nearer the second case of inefficient elimination of the core.

Consider first the global story. For a degenerate homogeneous body, the Virial theorem says that the intrinsic luminosity is approximately (Hubbard, 1968):

$$L = -d\,E_{thermal}/dt, \qquad (3)$$

where $E_{thermal}$ is only the thermal part of the internal energy content. This highly non-trivial (yet frequently assumed) result relies on the cancellation between the gravitational energy arising from contraction and the resulting work done on the material, primarily the positive internal energy increase arising from squeezing the electron gas.

It is easy to show that in the presence of redistribution of composition, the more complete result is approximately:

$$L = -d/dt(E_{thermal} + E_{diff}),$$

where $E_{diff}$ is the change in gravitational energy that arises from redistribution of the elements. It could represent the effect of helium transported down (where $E_{diff}$ becomes more negative as time progresses, thus adding to $L$) or, in our case, upward redistribution of rock and ice (where $E_{diff}$ becomes less negative as time progresses, thus decreasing $L$). More complicated cases can be envisaged, but the simplest interpretation of this is that the evolution cannot proceed (that is, convection and redistribution of anything will cease) unless $L > 0$ and accordingly $\Delta E_{thermal} > \Delta E_{diff}$, where $\Delta E_{thermal}$ is interpreted as the *decrease* in thermal energy of the planet over time and $\Delta E_{diff}$ is interpreted as the *increase* in gravitational energy arising from upward transport of heavies.

This statement is a necessary but not sufficient condition because of the First Law of thermodynamics. It is a global statement and it cannot be converted into an assertion about particular parts of the planet. It turns out to be similar (to order of magnitude) to the appealing notion that one must create buoyant material for the following reason: The scale height for the adiabatic distribution of temperature deep within the planet is $C_p/\alpha g \sim R$ (the planet radius), whence $\Delta E_{thermal} \sim M\alpha\Delta T g R$ where $\Delta T$ is the mean temperature drop in the evolution. The buoyancy requirement says that this must be at least as large as $\sim M(\Delta\rho/\rho)gR$, where $\Delta\rho$ is the change of density upon homogenizing the planet compositionally at fixed T, but this is indeed the estimate of $\Delta E_{diff}$ if we were to assume $\Delta\rho = \rho\alpha\Delta T$ (i.e compositional and thermal effects on density are equal). However, we now encounter the startling fact that $\Delta E_{thermal} > \Delta E_{diff}$ even in the extreme case where an initial core is homogenously redistributed throughout the overlying hydrogen-helium. The reason is simple: the thermal energy lost in the early part of Jupiter's history (of order the first hundred million years after the cessation of accretion) is very large in standard models. In other words, the thermal effect on density is larger than the compositional effect (averaged globally), essentially because the compositional effect is only a few percent on the planet mass. This is possible provided the planet begins sufficiently hot.

Specifically, $\Delta E_{thermal} \sim 4 \times 10^{42}(\Delta T/10^4 K)$ erg whereas $\Delta E_{diff} \sim 1.2 \times 10^{41}$ ($M_{core}/20 M_\oplus$) erg for realistic values of $C_p = 2 \times 10^8$ erg/g.K, and mean g = 3000 cm/s². A straightforward way of understanding this using the Virial theorem is to simply observe that an early Jupiter with a core is actually substantially larger in radius than a present-day homogeneous Jupiter.

Consider now the local picture (which will also illuminate why the global analysis is likely to be misleading.) For convection to operate at any radius, we require:

$$B_{thermal} = <v \delta \rho_{thermal}> >> B_{comp} = <v \delta \rho_{comp}>,$$

where $B$ is buoyancy flux, v is the convective velocity, $\delta \rho$ refers to the density anomalies for temperature and composition and <...> denotes an average over a surface at fixed equipotential surface. Physically, this says that rising fluid elements must be positively buoyant and falling fluid elements must be negatively buoyant, since work must be done. In passing it should be noted that this is automatically satisfied by double diffusive convection (LeConte & Chabrier 2012) though with the thermal $B$ being as much as ten times larger than the compositional $B$ because the diffusivity of heat is a hundred times larger than the diffusivity of composition.

Standard models of Jupiter would attribute the heat flux and therefore thermal buoyancy at small radius as arising only from the heat that escapes from the material *within* that radius. In other words, $B_{thermal}$ becomes small in precisely the place where it is needed to do work against gravity. This local picture says that the energy available to mix a core upward is the excess thermal energy in the core itself:

$$E_{thermal,local} = C_{p,core} M_{core} \Delta T \sim 1.2 \times 10^{40}(\Delta T/10^4 K)(M_{core}/20 \, M_\oplus) \; < \Delta E_{diff}$$

where $\Delta T$ must now be interpreted as the decrease in core temperature only, potentially larger than that of the overlying H-He. The high molecular weight and low specific heat (~$10^7$ erg/g K for rock or ice) is an important part of why this estimate of the thermal energy is so low, insufficient for efficiently mixing the core upward. The numbers suggest that a sufficiently hot core might be marginally capable of mixing itself upwards, but note that this material is much denser than the hydrogen despite its potentially very high temperature, arguing against efficient mixing.

The stark difference between global and local pictures is easy to understand. In the global picture, the planet cools from above and dense cold plumes sink into the interior. In the local interpretation of this, these plumes do work by dissipating turbulently as they proceed downwards and do not have the ability to help the upward transport of denser material, except in the limited sense of making the core *smaller* (i.e. erosion from the top but not dispersal). This is surely an oversimplification, but how much is difficult to assess without numerical simulations that cover multiple scale heights (Jupiter has of order ten temperature scale heights between the centermost hydrogen and the photosphere).

It is worth noting that was there any process that adds hydrogen to the heavy material, then this can greatly aid the mixing by shifting the location of erosion outward and by utilizing the substantially higher specific heat of the mixture. Indeed, double diffusive convection (e.g., Leconte & Chabrier, 2012) permits substantial upward mixing, from an initial state that is already substantially dispersed (a "dilute core" at time zero in our language), but even when starting from an initially compact, fluid core (e.g., Moll et al. 2017).

The details of the process of core erosion are still uncertain, and the erosion efficiency is even more in doubt now given that formation models predict that the innermost part of the planet is stable against convection. It is now even questionable whether primordial Jupiter is convective after formation. Several formation models have shown that it is more likely to be conductive (Cumming et al., 2018) or dominated by layered-convection, a less efficient type of convection (e.g., Wood et al., 2013). Therefore, the possibility of core erosion should be investigated in detail under these scenarios: layered convection or convective mixing at later times ($\sim 10^8 - 10^9$ years) where a larger fraction of the planet is expected to be convective (although the convective velocities at later times are lower, reducing the efficiency of mixing). In addition, if core erosion has occurred it means that the deep interior should be convective, unless the redistribution of the material is via layered convection or one assumes that convective mixing led to the formation of steep enough composition gradients that convection was inhibited after a period of efficient mixing. Finally, another important property that needs to be determined in order to assess the likelihood of core erosion is the solubility of the expected core material in metallic hydrogen. DFT calculations suggest that $H_2O$, $M_gO$, and Fe are soluble in the deep interior of Jupiter which makes the possibility of core erosion more feasible (Wilson & Militzer, 2012a,b; Wahl et al. 2013). Nevertheless, more comprehensive calculations of convective mixing are still missing.

The arguments presented above are based on fundamental principles but somewhat simplified considerations. They suggest that the erosion of a heavy-element core in young Jupiter might be possible. It is clearly required to model core erosion using numerical models with more realistic conditions. These models, however, while they might be more realistic, still use parameters that are poorly known under Jovian conditions such as the mixing parameter (i.e., the ratio of the mixing length to the scale height), the number of layers in the case of layered convection, as well as the opacity and the diffusivities. Only with significant progress in experiments, which could constrain these values, will the numerical models be more representative of the reality.

## 4.2. Evolution models with convective mixing

In order to link Jupiter's present internal structure with its origin, it is crucial to understand the planet's long-term evolution. The internal structure can change with time due to various physical and chemical processes such as convective mixing, settling, phase separation, etc. By modeling the evolution of Jupiter, we can identify what primordial internal structure can lead to its internal structure today, and in this way to better understand its formation. While this is a challenging task, efforts in this direction are ongoing.

Vazan et al. (2018) presented Jupiter's evolution with an extended primordial composition gradient. This model assumes the presence of a compositional gradient immediately after accretion but before the subsequent thermal evolution. It was found that the outer part of the composition gradient, corresponding to nearly 50% of the planetary radius, becomes homogenous via convective mixing after several million years and that this mixing leads to an enrichment of the planetary envelope with heavy elements, while the deep interior remains stable against convection due to the steep gradient. This affects the thermal evolution and leads to hotter interiors in comparison to the standard adiabatic case. As in the structure models with layered convection, the total heavy element mass in the planet is higher than in the adiabatic models and was found to be about 40 $M_\oplus$.

In principle, the atmospheric enrichment (perhaps a factor of three over solar, not a small effect) could come either from late addition of material or core erosion. It should be noted that the observed enrichment in Jupiter's envelope includes heavy noble gases. In the usual way of thinking about the formation, the primordial core does not include these noble gases because that material may have formed at a location where T ~100K or even more. This would be needed to explain the absence of large amounts of water ice inward of that location. Under those conditions, incorporation of large amounts of heavy noble gases is not expected, since this is thought to be a low temperature process relative to T~100 K (perhaps as low as 40K). Still, it must be admitted that our understanding of the migration of bodies in the early solar system is still imperfectly understood. It is commonly supposed that most or all of the enrichment of Jupiter's envelope is not a consequence of upward mixing from the core but is instead caused by late accretion of material, planetesimals but potentially also gas (see section 6), that was derived from lower temperature regions of the disk (e.g., Owen et al., 1992; Guillot & Hueso, 2006; Mousis et al., 2019; Oberg & Wordsworth, 2019; Bosman et al. 2019) in a process that is typically not included in the standard formation models.

A direct consequence from the work of Vazan et al. (2018) is to show that a situation in which a composition gradient is maintained throughout the giant planet's envelope, as envisioned by Leconte & Chabrier (2012) becomes rapidly unstable to convection. Cooling from the top implies that the envelope becomes rapidly mixed from the outside in, erasing any compositional gradient at least in a large outer part (50% or more) of the envelope, with the notable exception of zones in which phase separation occur, such as the hydrogen-helium phase separation (see Mankovich et al. 2016). Another consequence is that this mixing of high-Z material from below leads to an enrichment of the atmosphere which may potentially explain the fact that giant planets are enriched compared to the protosolar value (see Atreya et al. 2019).

The Vazan et al. (2018) model, like many other published interior structure models, does not attempt to realistically simulate formation model of the planet. The initial temperature and composition profile were chosen ad-hoc so that the final structure at 4.5 Gyrs matches observations. As a result, the initial model is significantly colder than the one found by formation models that properly model the planetary growth and the accretion shock during phase-3.

An attempt to more realistically link Jupiter's origin with its thermal evolution was recently presented by Müller et al. (2020). In this study Jupiter's formation and long-term evolution were modeled from the onset of runaway gas accretion until today, accounting for the energy transport and heavy-element mixing. The models were constructed using different formation scenarios, with primordial composition gradients, as well as with various heavy-element accretion rates and shock properties during runaway gas accretion. The thermal properties of the forming planet depend on the radiative efficiency of the accretion shock, which under some conditions can lead to an extended radiative envelope (Cumming et al., 2018).

After Jupiter's mass is reached, runaway gas accretion terminates and Jupiter's long-term evolution to the present-day is modeled accounting for the mixing of heavy elements in the interior. In Müller et al. (2020) it was assessed whether primordial composition gradients in Jupiter's interior can be sustained. They found that in all the models they considered after a few tens of millions of years the envelope becomes homogenous, with the dilute core being limited to the innermost 20% of the planetary mass. These models that properly account for the planetary growth and the heating associated with gas accretion are much hotter throughout most of the envelope than the one used by Vazan et al. (2018) and the heavy-element gradient is insufficient to inhibit large-scale convection unless proto-Jupiter is unrealistically cold. In addition, in order to form extended primordial composition gradients an extremely large mass of heavy elements must be accreted during runway gas accretion, which is also rather unrealistic

(Shibata et al., 2019). Therefore, more realistic formation and evolution models imply that composition gradients might be hard to sustain on a timescale of Gyr, unless Jupiter had formed much colder than predicted from current formation models (Müller et al., 2020, Valetta & Helled, 2020).

One might conclude that explaining Jupiter's dilute core is challenging for standard formation models. As we discuss above, the dilute core could be an outcome of a different formation history. At the moment, there is not yet a coupled formation-evolution model that can explain Jupiter's internal structure. Finding a self-consistent formation and evolutionary path that leads to Jupiter's internal structure today as predicted by structure models is certainly desirable but has yet to be achieved.

## 5. Summary & Outlook

The Juno mission has provided new and exciting measurements of Jupiter that are now being applied to better understand Jupiter's formation, evolution, and interior. A key result from interior models that fit Juno's accurate gravity data is that the planet is not fully mixed with a heavy-element mass fraction that increases towards the center but instead consists of a "dilute" or "fuzzy" core which extends up to as much as half of Jupiter's radius. The excess heavy element mass in that region is still being investigated.

As discussed here, standard core accretion formation models that account for the accretion of the heavy elements during the planetary growth naturally predict composition gradients in the deep interior that could mimic a fuzzy core. If the growth time of the planet is short, and runaway gas accretion occurs when the protoplanet has a mass of $\sim 30 M_\oplus$ the region containing composition gradients is significantly smaller than the one predicted from structure models (e.g., Helled & Stevenson, 2017, Bodenheimer et al., 2018, Müller et al., 2020). However, an extended dilute core with a heavy-element enrichment of $Z \sim 0.3$ can be formed if the planet has a long phase-2, dominated by planetesimal accretion that provides sufficient energy to delay gas accretion (Alibert et al., 2018, Venturini & Helled, 2019, Valletta & Helled, 2020). In that case, runaway gas accretion can be delayed until a mass of about 100 $M_\oplus$ compatible with Jupiter's structure models. Alternatively, Jupiter's dilute core could be a result of a head-on giant impact of a massive impactor diluting a primordial compact core (Lui et al., 2019).

Juno data have clearly revolutionized our understanding of giant planets, and the extended mission promises more data. For example, connecting the gravity and magnetic fields is important: A region with a dilute stable core with a radius above 0.5 $R_J$ not only introduces challenges to planet formation theory but also has implications for the generation of Jupiter's magnetic field. sThe presence of static stability inferred for Saturn from seismology (Mankovich et al. 2021) may indirectly help us understand Jupiter's internal structure. The tidal response of Jupiter due to Io has been measured and might tell us about the static stability ("dilute core"). Measurement of the precession constant could test our understanding of the moment of inertia, currently predicted to be tightly constrained by the measured gravity. Continue refinement of interior models coupled with additional microwave measurements will better constrain the water abundance and further narrow the ideas for how Jupiter acquired its peculiarly Z-enriched envelope. In the long term, seismology (presumably Doppler imaging) may prove to be key to understanding Jupiter's structure. In the meantime, improved knowledge of the EoS of the different elements and their interaction, are crucial in combination with all the available information (gravity field, magnetic field, atmospheric composition, etc.) to further constrain Jupiter's internal structure.

Since Jupiter is our local ground truth for giant planets, the insights into its formation and evolution are applicable to giant exoplanets, with the aim to better understand their formation and evolution and put our Solar System in perspective. Although many questions remain open, this is a golden era in giant planet exploration. The measurements of the ongoing extended Juno mission, the characterization of many giant planets around other stars, and the theoretical efforts provide new insights on the nature of gas giant planets.

**ACKNOWLEDGEMENTS**

RH thanks S. Müller, S-F. Lui, A. Vazan & C. Valletta as well as support from the Swiss National Science Foundation (SNSF) via grant 200020 188460. We thank the two anonymous referees for valuable comments. JIL gratefully acknowledges support from the Juno mission through the Southwest Research Institute, Contract number NNM06AA75C.